\documentclass[aps,twocolumn,showpacs,amsmath,amssymb,pra,superscriptaddress,floatfix,longbibliography]{revtex4-1}

\usepackage{times}
\usepackage{t1enc}
\usepackage{subfigure}
\usepackage{graphicx}
\usepackage{epstopdf}
\usepackage{amsmath}
\usepackage{dcolumn}
\usepackage{color} 
\usepackage{soul}

\begin{document}


\title{Atom recoil in collectively interacting dipoles using quantized vibrational states}

\author{Deepak A.~Suresh}
\affiliation{Department of Physics and Astronomy, Purdue University, West Lafayette,
Indiana 47907, USA}

\author{F.~Robicheaux}
\email{robichf@purdue.edu}
\affiliation{Department of Physics and Astronomy, Purdue University, West Lafayette,
Indiana 47907, USA}
\affiliation{Purdue Quantum Science and Engineering Institute, Purdue
University, West Lafayette, Indiana 47907, USA}




\date{\today}

\begin{abstract}
The recoil of atoms in dense ensembles during light matter interactions is studied using quantized vibrational states for the atomic motion. The recoil resulting from the forces due to the near-field collective dipole interactions and far-field laser and decay interactions are explored. The contributions to the recoil and the dependence on the trap frequency of the different terms of the Hamiltonian and Lindbladian are studied. These calculations are compared with previous results using the impulse model in the slow oscillation approximation. Calculations in highly subradiant systems show enhanced recoil indicating that recoil effects cannot be ignored in such cases. 

\end{abstract}


\maketitle

\section{Introduction\label{sec:intro}}


The study of collective dipole-dipole interactions has progressed significantly since Dicke pioneered the idea in 1954 \cite{dicke,bettles,re1971,fhm1973,gfp1976,s2009,mss2014,pbj2014,bbl2016,bzb2016,psr2017,jbs2018}. There have been many recent innovations using collective interactions in coherent quantum control and quantum information \cite{zoller,chs2003,ymg2013,mmm2018,jrp2017,bpp2020,ggv2018,wzc2015}. As an example, atom arrays which are densely packed have been shown to have high reflectivity \cite{bga2016,swl2017,bloch}. But as the atoms get closer and denser, the forces due to the collective interactions become larger causing the internal states to become entangled with the vibrational motion of the atom. This causes unwanted decoherences to arise in the system.
There is a need to better understand the forces involved with the collective dipole interactions and the role that recoil plays in the coherence of the system. 


These questions motivated Refs. \cite{shihua,sr2021}, where we studied the recoil in the atoms in light matter interactions in densely packed ensembles and atom arrays. More specifically, in Ref \cite{sr2021}, we described a model to calculate the recoil in atoms where the photon recoil is considered as an impulsive force. This model will be referred to as the impulse model in this paper. 
It was constructed under the slow oscillation, or equivalently, the sudden approximation, where the timescales of the atomic oscillations are much longer than the timescales of the internal state dynamics.
This implied that the trap frequencies should be much smaller than the decay rate of the system. 

Typically, the trap frequencies used are 10 to 100 kHz while the decay rates of electronic excitations are often around  10s of MHz. While these trap frequency ranges would normally be within the sudden approximation, problems arise when the system becomes subradiant and the collective decay rates approach the trap frequencies.
The results from the impulse model also indicated that the recoil is typically proportional to the lifetime of the excitation in certain cases, leading to enormous recoils in highly subradiant systems. While the sudden approximation gives an intuitive understanding of how energy is added to the center of mass motion and how decoherence arises, the assumptions in the approximation are dubious for some of the more interesting atomic arrangements. The goal of this paper is to clarify such ambiguous results and to extend the analysis beyond the approximations used in Ref. \cite{sr2021}.

The quantum harmonic oscillator model described in this paper, calculates the recoil in collectively interacting systems  but removes the assumptions in the sudden approximation. The $N$ atoms are assumed to be trapped in harmonic potentials having quantized vibrational energy states. Using the density matrix formalism, we time evolve the combined-vibrational and internal state density matrix, to calculate the momentum and energy deposited in the system at a later time. 
This model does not have the limitation of the sudden approximation and can be used to simulate a wide range of trap frequencies and, thus, can serve as an important test of the sudden approximation. It will also provide insight into how the different terms of the Hamiltonian and the Lindblad operator contribute to the recoil of the atoms. We focus on the transfer of energy in the system rather than the vibrational population as the population in the excited states trivially decrease as the frequency increases for the same energy transfer.

To simplify the calculations, we will work in the low intensity limit where there is only a single excitation in the system, i.e., only one atom can be electronically excited at a time. This will reduce the number of internal states from $2^N$ to $(N+1)$. We also only investigate cases where the spread of the atomic wavefunction is smaller than the distances of atom separation, to reduce the overlap of wavefunctions. This is expected in reasonable experimental arrangements because otherwise the atom grid isn't well defined.

This paper proceeds as follows. Section \ref{sec:methods} discusses the model and equations used. Section \ref{sec:singleatom} discusses the decay and laser interaction for a single atom to illustrate the role of recoil and Sec. \ref{sec:twoatoms} extends the analysis to $N$ atoms. We discuss approximations to simulate a large  number of atoms in Sec. \ref{sec:approximations} to calculate the recoil in arrays of atoms and subradiant cavities. Section \ref{sec:conclusion} presents the conclusions and summarises the results and future outlook.

\section{Methods\label{sec:methods}}
We shall consider $N$ atoms, each trapped in a quantum harmonic potential with each atom having two internal electronic states. The center of each trap will form a spatial arrangement required by the experiment, for example, a square array. Since the atoms are in a harmonic trap potential, they will each have an infinite Hilbert space of vibrational levels.  
We can limit the number of vibrationally excited states for each atom to be states $n < N_{vib}$ for calculation purposes. 
When the spread of the atomic wavefunction is small, the effects on the harmonic oscillator wavefunctions are separable across the different directions. Hence, we can run the calculations by choosing one oscillation direction at a time. 
The N atoms together will have a combined vibrational Hilbert space of $V = (N_{vib})^N$ states. Since we are working in the low intensity limit and only one atom can be electronically excited at any time, the total number of internal states is N + 1. Hence, the total number of states is $(N + 1)(N_{vib})^N$. 

The internal states will be represented by $| j \rangle$, the collective vibrational states will be represented by $| m \rangle$ and the total state will be denoted by $| j, m \rangle$. The internal state index goes from 0 to N, where $j = 0$ represent the electronic ground state (alternatively $| g \rangle$) with no atom excited and $j = 1$ to $N$ represent only the jth atom being excited. The collective vibration state $| m \rangle$  is the tensor product of all  possible vibrational states, i.e, $| m \rangle = | n_1 \rangle \otimes | n_2 \rangle \otimes  … \otimes | n_N \rangle$ where $| n_i \rangle$ is the vibrational state of the ith atom. The index $m$ goes from $m = 0$ to $(N_{vib})^N - 1$.

Hence, the density matrix will be represented by 
\begin{equation}
    \rho = \sum_{j,j'}\sum_{m,m'} \rho_{j,j'}^{m,m'} | j,m \rangle \langle j',m'|
    \label{eqn:rho}
\end{equation}

The density matrix evolves according to the equation given by 
\begin{equation}
    \frac{d\hat{\rho}}{dt} = -\frac{i}{\hbar}[\hat{H}, \hat{\rho}] + \mathcal{L} (\hat{\rho})
    \label{eqn:rhodot}
\end{equation}
where $\hat{\rho}$ is the density matrix of the system, $\hat{H}$ is the effective Hamiltonian and $\mathcal{L}(\hat{\rho})$ is the Lindblad super-operator. 
The effective Hamiltonian consists of three parts. (1) The trap potential of the atoms, which is a quantum harmonic oscillator Hamiltonian, (2) the laser Hamiltonian, and the (3) dipole-dipole resonant interaction.

The Hamiltonian of the trap potential will be given by
\begin{equation}
    H_t = \sum_j \hbar \omega_t ( a^{\dagger}_j a_j + \frac{1}{2}) 
    \label{eqn:Htrap}
\end{equation}
where $\omega_t$ is the trap frequency of the harmonic oscillator and $a^{\dagger}_j$ and $a_j$ are the harmonic oscillator ladder operators for the $jth$ atom in the chosen direction. 
The mean position of the wavefunction or the fixed point positions of the atoms will be given by $\mathbf{R}_j$ and the the spread of the atom or the position of the atom with respect to the mean will be given by $\mathbf{r}_j$. The position operator along the chosen direction is given by $s_j = \sqrt{\frac{\hbar}{2M\omega_t}} (a_j + a_j^\dagger)$. We define the quantity $\kappa = k\sqrt{\frac{\hbar}{2M\omega_t}}$ where the length-scale for the atoms' motion and  the spread of the atomic wavefunction is described by $\kappa / k$. Here, $k$ is the wavenumber of the resonant light and $M$ is the mass of a single atom.

When the laser interacts with the atoms, it imparts a momentum of $\hbar k$ which will manifest in the Hamiltonian through the position operators $s_j$
The Hamiltonian due to the laser is
\begin{equation}
\begin{split}
    \hat{H}_L =  \hbar \sum_j \left[ -\delta\hat{\sigma}_j^+\hat{\sigma}_j^- + \bigg( \frac{\Omega }{2}\hat{\sigma}_j^+ e^{i \mathbf{k}_0\cdot (\mathbf{R}_j + \mathbf{\hat{r}}_j)}  + h.c.\bigg) \right]
    \label{eqn:Hlaser}
    \end{split}
\end{equation}
where $\Omega$ is the Rabi frequency, $\delta$ is the detuning and $\mathbf{k}_0 = k \mathbf{\hat{z}}$ as the initial wavevector of the incoming photons. $\hat{\sigma}_j^+$ and $\hat{\sigma}_j^-$  are the raising and lowering operator of the electronic excitation of the jth atom.
If the laser is propagating in the chosen direction of vibrational oscillation, the term $\mathbf{k}_0\cdot\mathbf{r}_j $ can be replaced by $\kappa (\hat{a}_j + \hat{a}_j^\dagger)$. Otherwise, the $\mathbf{k}_0\cdot\mathbf{r}_j $ term will be dropped and the laser will not cause any vibrational transitions.

In the following equations, the primed and unprimed coordinates are used to signify either a right or left multiplication of the density operator respectively. For signifying differences, we will use the following convention
\begin{equation}
    \mathbf{r}_{ij} \equiv \mathbf{r}_i - \mathbf{r}_j;   \qquad 
    \mathbf{r}_{ij}' \equiv \mathbf{r}_i - \mathbf{r}_j';  \qquad   
    \mathbf{r}_{ij}'' \equiv \mathbf{r}_i' - \mathbf{r}_j'
    \label{eqn:rprime}
\end{equation}
The resonant dipole-dipole interactions are given by the imaginary part of the Lindblad term and is given by
\begin{equation}
    \hat{H}_{dd} = \hbar \sum_{i \neq j} Im\{g(\mathbf{R}_{ij}+\mathbf{r}_{ij}) \}\hat{\sigma}_i^+ \hat{\sigma}_j^-
    \label{eqn:Hdd}
\end{equation}
The real part of the Lindblad term describes the dynamics of the decay and is given by
\begin{equation}
\begin{split}
    &\mathcal{L}(\hat{\rho}) = \sum_{i,j}\big[
    2Re\{g(\mathbf{R}_{ij} +\mathbf{r}_{ij}')\}\hat{\sigma}_i^- \hat{\rho} \hat{\sigma}_j^+  \\
    &- Re\{g(\mathbf{R}_{ij}+\mathbf{r}_{ij})\}\hat{\sigma}_i^+ \hat{\sigma}_j^- \hat{\rho} 
    - \hat{\rho}\hat{\sigma}_i^+ \hat{\sigma}_j^-Re\{g^*(\mathbf{R}_{ij}+\mathbf{r}_{ij}'')\}
    \big]
    \label{eqn:lindblad}
\end{split}
\end{equation}
where the Green's function $g(\mathbf{R})$ is given by
\begin{equation}
\begin{split}
     g(\mathbf{R}) = \frac{\Gamma}{2}&\bigg[h_0^{(1)}( kR)
     +\frac{3(\hat{R}\cdot\hat{q})(\hat{R}\cdot\hat{q}^*) - 1}{2} h_2^{(1)}(kR)
     \bigg]
     \label{eqn:greens}
\end{split}
\end{equation}
where $\hat{q}$ is the dipole orientation, $R = |\mathbf{R} |$ is the norm of $\mathbf{R}$, $\hat{R} = \mathbf{R}/R$ is the unit vector along $\mathbf{R}$, $\Gamma$ is the decay rate of a single atom and $h_l^{(1)}(x)$ are the outgoing spherical Hankel function of angular momentum $l$; $h_0^{(1)}(x)=e^{ix}/[ix]$ and $h_2^{(1)}(x) = (-3i/x^3 - 3/x^2 + i/x)e^{ix}$.

When we calculate the Green's function, we take a Taylor expansion up to second order which will be valid under the condition that the spread of the wavefunction $(\kappa/k) \ll $ the separation of atoms.

\begin{equation}
\begin{split}
    g(\mathbf{R}_{ij} + \epsilon) = g(\mathbf{R}_{ij}) + \big(\frac{g'(\mathbf{R}_{ij})}{k}\big) k\epsilon + \big(\frac{g''(\mathbf{R}_{ij})}{k^2}\big) \frac{k^2\epsilon^2}{2} + ...
     \label{eqn:g_taylor}
\end{split}
\end{equation}
where the derivatives are taken in the chosen oscillation direction. Since the Hankel functions in $g(R)$ are functions of $kR$, the $k$'s in the denominator make the expansion term, $k\epsilon$, more explicit. The $\epsilon = s_i - s_j$ is expanded into the corresponding vibrational ladder operators.
The zeroth order term does not depend on the spread of the atoms and does not cause any transitions in the vibrational state. The first and second order terms depend on the spread of the wavefunction and will induce single level and two level transitions in the vibrational states respectively. 

Since the Green's function depends on both $\mathbf{r}_j$ and $\mathbf{r}_j'$, which correspond to the left or right operation on the density matrix, we have $s_j$ and $s_j'$ respectively.  While the last two terms of the Lindblad expression only have left or right multiplication, the first term behaves differently. Upon expanding the Harmonic vibrational wavefunctions, we see that the first term acts on the combined density matrix as

\begin{equation}
    \begin{split}
        &Re\{g(R_{ij} + r_{ij}') \} \hat{\sigma}_i^- \hat{\rho} \hat{\sigma}_j^+   =  Re\{g(R_{ij}) \} \hat{\sigma}_i^- \hat{\rho} \hat{\sigma}_j^+\\
        &+  Re\{\frac{g'(R_{ij})}{k} \} k(s_i\hat{\sigma}_i^- \hat{\rho} \hat{\sigma}_j^+ - \hat{\sigma}_i^- \hat{\rho} \hat{\sigma}_j^+ s_j')\\
        &+ Re\{\frac{g''(R_{ij})}{k^2} \} \frac{k^2}{2} ( s_i^2 \hat{\sigma}_i^- \hat{\rho} \hat{\sigma}_j^+ + \hat{\sigma}_i^- \hat{\rho} \hat{\sigma}_j^+ s_j'^2  - 2 s_i \hat{\sigma}_i^- \hat{\rho} \hat{\sigma}_j^+ s_j')
        \label{eqn:ReG}
    \end{split}
\end{equation}
where the $ks_i$'s will be replaced by $\kappa ( a_i + a_i^\dagger)$ notation when solving the equations.
The expectation values of the momentum and energy in the vibrational state of atom $j$ can then be calculated from the density matrix, 
\begin{equation}
   p_j = \frac{i}{2\kappa} Tr\left[  (a_j^{\dagger} - a_j ) \rho\right] \hbar k
    \label{eqn:OscMomentum}
\end{equation}
\begin{equation}
    E_j= \frac{1}{2\kappa^2} Tr\left[  (2a_j^{\dagger}a_j + 1 ) \rho\right] E_r
    \label{eqn:OscEnergy}
\end{equation}
where $\hbar k$ and $E_r = \hbar^2 k^2/(2M)$ are the recoil momentum and energy deposited when one photon is absorbed or emitted by an atom. Since the expression for the energy is divided by $\kappa^2$, only the terms of the order $\kappa^2$ in the diagonal of the density matrix will primarily contribute to a change in energy. The contribution from the $\kappa^4$ terms and beyond will be negligible for small wavefunction spreads. The energy difference in the vibrational levels will be given by $1/\kappa^2$. That is, if $\kappa = 0.01$, the energy difference of consecutive vibrational levels will be $10^4 E_r$. 
\section{Results}

The quantum oscillator model can be used to study the forces involved in the collective dipole interactions at high atom densities. We can study the contribution from the different terms of the Hamiltonian and the Lindblad equations. Since this model does not have any restrictions with respect to the trap frequency, we can investigate the validity of the sudden approximation.

We are more interested in the cases with higher trap frequencies where the vibrational energy spacing is much larger than $E_r$. Hence we do not need to include many vibrational levels. This also implies that the spread of the wavefunction will be small and we can limit the Taylor series expansion, Eq. \eqref{eqn:g_taylor}, to second order terms. 



\subsection{Single atom decay\label{sec:singleatom}}
To begin, we analyse the simple decay process of a single atom trapped in a harmonic potential. The atom is initially excited and no laser interaction is present. The effective Hamiltonian becomes
\begin{equation}
    H_t = \hbar \omega_t ( a^{\dagger}_1 a_1 + 1/2) 
    \label{eqn:SA_Htrap}
\end{equation}
Since the $H_t$ is purely diagonal with respect to the vibrational states, its contribution to the change in the density matrix, 
\begin{equation}
    \Dot{\rho} = \frac{-i}{\hbar} [H_t,\rho] 
    \label{eqn:SA_Hrhodot}
\end{equation}
has zero diagonal elements and only interacts with the off-diagonal coherence terms of the density matrix. The Lindblad term for a single atom is
\begin{equation}
\begin{split}
    &\mathcal{L}^{(1)}(\rho) = 2 Re\{g(r_{11}')\}\sigma_1^- \rho \sigma_1^+\\
    &-Re\{g(r_{11})\}\sigma_1^-  \sigma_1^+\rho
    -\rho\sigma_1^-  \sigma_1^+ Re\{g(r_{11}'')\}
    \label{eqn:SA_lindblad}
\end{split}
\end{equation}
Since $ r_{11} = r_{11}'' = 0$, $Re\{g(r_{11})\} = Re\{g(r_{11}'')\} = \Gamma / 2$. The last two terms do not contribute to change in the vibrational states. Expanding the first term using $r_{11}' = 0 + s_1 - s_1'$ up to the second order gives
\begin{equation}
    Re\{g(r_{11}')\} = g(0) + \frac{g''(0)}{2} \kappa^2 (s_1-s_1')^2
    \label{eqn:SA_ReG}
\end{equation}
since the first derivative $g'(0) = 0$. The next non-zero leading order term will be the fourth order, since the third derivative is again 0, but they will be of the order $\kappa^4$ and will not cause significant contributions when calculating the energy.
If we assume the atom is initially excited and in the vibrational ground state, the first term of Eq. \eqref{eqn:SA_lindblad} becomes
\begin{equation}
\begin{split}
    \mathcal{L}_1(\rho) =  | g \rangle \langle g | \rho_{1,1}^{0,0} \bigg(\big(\Gamma &+ 2 \kappa^2g''(0)\big) \, |0\rangle \langle 0|\\ 
    &- 2  \kappa^2g''(0)\,\,\, |1\rangle \langle 1|\\
    + \kappa^2 & g''(0) \sqrt{2} (|2\rangle \langle 0| + |0\rangle \langle 2|)\bigg)
    \label{eqn:SA_L1}
\end{split}
\end{equation}

We can analytically solve the above equation to obtain the change in the vibrational energy at infinite time when the decay is complete. The change in vibrational energy is given by
\begin{equation}
    \Delta E = -2 \frac{g''(0)}{\Gamma} E_r
    \label{eqn:SA_Energy}
\end{equation}
This result remains valid when the initial density matrix is any incoherent combination of vibrational states. For an atom initially excited and polarized in the $e_+ = - ( \mathbf{\hat{x}} + i \mathbf{\hat{y}} ) / \sqrt{2}$ direction, $\Delta E_z = 0.4 E_r$ and $\Delta E_x =\Delta E_y = 0.3 E_r$.
The energy deposited due to the recoil from the emission of a single photon is independent of the frequency of the harmonic oscillator. 
This result is correct even if we go beyond the second order approximation in Eq. \eqref{eqn:SA_ReG}.

\subsubsection{Laser interaction\label{sec:laser}}


When the atoms absorbs a single photon from the laser, there is a momentum of $\hbar k$ added to the atoms. The contribution to the change in vibrational state comes as $e^{iks_1}$ in Eq. \eqref{eqn:Hlaser}. Since $\kappa$ is small, a Taylor expansion gives
\begin{equation}
\begin{split}
    e^{iks_1} &= 1 + i k s_1 - \frac{k^2s_1^2}{2}+...\\
    & = 1 + i \kappa (a^{\dagger}_1 + a_1)  - \frac{\kappa^2}{2} (a^{\dagger}_1 + a_1)^2 + ...
    \label{eqn:SA_laser}
\end{split}
\end{equation}
Since the laser interacts with the density matrix through the coherence terms, the order of the transitions to the population from the first order and second order terms are $\kappa^2$ and $\kappa^4$ respectively. Hence, the energy deposited is primarily contributed by the first order term.

\begin{figure}
    \centering
    \includegraphics[width=0.45\textwidth]{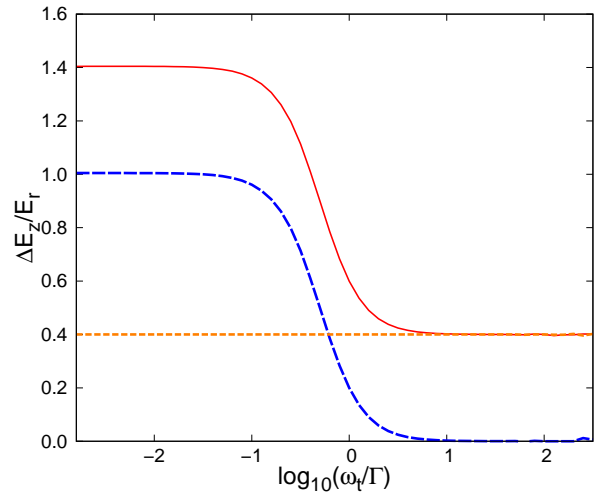}
    \caption{The vibrational energy deposited, per incoming photon, at steady state for a laser incident on a single atom. The red solid line shows the total energy deposited while the blue dashed and orange dashed lines show the contribution from the coherent laser transfer and the decoherent decay. The calculations were run using $N_{vib} = 5$.}
    \label{fig:1At}
\end{figure}

When there is a continuous laser incident on the atoms, the electronic internal states of the atoms reach a steady state. 
Instead of the total recoil energy and momentum deposited, we calculate the rate of recoil deposited in the atoms by time evolving the density matrix using Eq. \eqref{eqn:rhodot}. 
Figure \ref{fig:1At} shows the energy deposited per incident photon in the direction of the incident laser on a single atom as we vary the trap frequency. It also shows the contribution of the kick due the coherent laser interaction and the decoherent single atom decay term. To ignore long term effects like shifts in position due to radiation pressure, the expectation values are taken immediately after reaching electronic steady state. It is important to note that we are discussing the transfer of energy across different trap frequencies and not the population in the excited states. As the frequency goes up, the energy difference between the vibrational states will increase. If the energy transfer remains the same but the frequency goes up, there will necessarily be less population in excited vibrational states.

The atom absorbs a photon and randomly emits it in an arbitrary direction. At low trap frequencies, the absorption of the photon results in $E_r$ recoil and the emission gives $0.4 E_r$ in the laser direction. The recoil due to the emission agrees with the result of Eq. \eqref{eqn:SA_Energy}, and is independent of the trap frequency.
But as we increase the frequency, the contribution for vibrational excitation from the laser becomes negligible. 
At low trap frequencies ($\omega_t \ll \Gamma$), the vibrational energy states are close enough that the linewidth spread of the excited state 
can allow vibrational transitions. On the other hand, at high trap frequencies ($\omega_t \gg \Gamma$), the vibrational energy states are far enough apart that there is no vibrational transitions due to the laser. 
Hence the kick from lasers reduce when the trap frequency is higher than the decay rate of the system. Effects such as side-band cooling can also be seen when the trap frequencies are higher than the decay rate.

\subsubsection{Coherent and decoherent transfers\label{sec:coherent}}
There are two types of vibrational populations transfer occurring in the system. When the {\it population} transfers through the coherence terms (off-diagonal terms) of the density matrix it is called coherent transfer. This is a two-step process where the initial population terms couple to coherence terms which then couple to population terms in different vibrational states ultimately leading to a change in vibrational energy. Hence, any coherent transfers of the order $\kappa^2$ will lead to a population change of the order $\kappa^4$. The transfers due to the laser Hamiltonian are an example.

If the population directly transfers between the diagonal terms, without going through the coherence terms, they are called decoherent transfers. This can be seen in the second line of Eq. \eqref{eqn:SA_L1}, where there is a direct single level transition from the $|0\rangle \langle 0|$ to $|1\rangle \langle 1|$ vibrational state. 
Since the trap Hamiltonian only acts on the coherence terms, they do not affect the dynamics of the decoherent transfers. Hence the decoherent {\it energy} transfers, such as the single atom decay term, are unaffected by the trap frequency. 




\subsection{Multi-atom decay\label{sec:twoatoms}}

When there is more than one atom interacting, the $H_{dd}$ Hamiltonian [Eq. \eqref{eqn:Hdd}] and the two atom Lindblad terms [i.e, $i \neq j$ terms of Eq. \eqref{eqn:lindblad}] come into effect.
Since the vibrational raising and lowering operators in these terms act on different atoms, they cannot directly transfer vibrational population. They go through the coherence terms and are coherent population transfers, see Sec. \ref{sec:coherent}.

\begin{figure}
    \centering
    \includegraphics[width=0.45\textwidth]{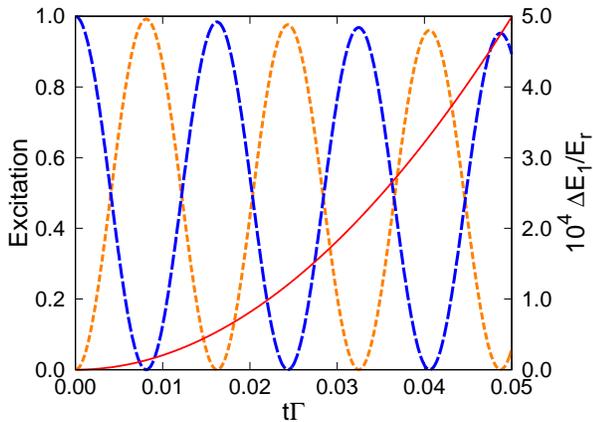}
    \caption{The excitation is exchanged between two atoms which are very close to each other ($d = 0.02 \lambda$) when one atom is initially excited. Orange and blue dashed lines indicate the excitation probability of the two atoms. Red line shows the increase in the vibrational energy of the first atom. The calculation was done using $\kappa = 0.00001$ and $N_{vib} = 2$ using the full density matrix.}
    \label{fig:2At}
\end{figure}

For simplicity, we can look at the case of two atoms. When there are two atoms very close to each other $(d \approx 0.02 \lambda)$, and one of the atoms is excited, the excitation rapidly hops between the two atoms while decaying, as seen from Fig. \ref{fig:2At}. This is the resonant dipole-dipole interaction arising from the Hamiltonian term from Eq. \eqref{eqn:Hdd}. Even though the excitation probability of the atom alternates, the recoil energy deposited on the atom increases continuously. All the recoil in this timescale comes from the near-field dipole dipole interactions, i.e., through the two atom dipole-dipole Hamiltonian [Eq. \eqref{eqn:Hdd}]. 

When two atoms interact, the direction along the line connecting the atoms and the directions perpendicular have considerably different physics. Let the atoms be separated in the x-direction by a distance $ d < \lambda$. In the directions along the separation, ie, in the x direction,  there are interatom forces that arise due to the collective interactions. These forces act only along the line joining the two atoms. 
In the directions perpendicular to the separation, i.e., y and z-direction, there are no interatom forces and only the kick from the photon emitted contributes to the recoil. 

\subsubsection{Transverse Oscillation}

For two atoms, when the chosen direction of vibrational quantization is perpendicular to the separation of the atoms, there are no inter-atom forces. 
While taking the Taylor expansion, the first derivative of the Green's function $g'(\mathbf{R}_{ij})$ in the direction perpendicular to the separation is zero.
This results in the equations being similar to the equations for the single atom case, where only zero and second order terms remain. But since the two atom Lindblad terms are coherent transfers, the second order term of $\kappa^2$ will contribute to only a $\kappa^4$ order of vibrational population transfer. Hence we see that in the perpendicular direction, only the contribution from the single atom Lindblad terms contribute to the change in vibrational energy to the lowest order in $\kappa$. The single atom terms being decoherent transfers also implies that the energy deposited in the perpendicular direction is independent of the trap frequency. Thus, the impulse model is valid even beyond the sudden approximation in the directions where there are no inter-atom interactions i.e., perpendicular to the atom array. 

\begin{figure}
    \centering
    \includegraphics[width=0.45\textwidth]{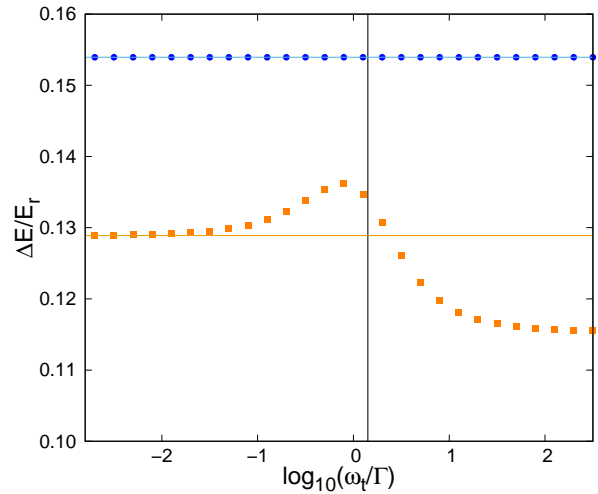}
    \caption{The energy deposited during the decay of two atoms uniformly excited, separated by $d = 0.4 \lambda$ in the x-direction, verses the trap frequency. The blue circles and orange squares indicate the quantum harmonic oscillator model results in the z and x-direction respectively. The thin solid lines indicate the respective impulse model result. The black vertical line denotes the collective decay rate of the system. The calculations are done using full density matrix with $\kappa = 0.001$ kept as constant while varying $w_t$.}
    \label{fig:2Atdecay}
\end{figure}

Figure \ref{fig:2Atdecay} shows that the recoil in the perpendicular direction is independent of the frequency and agrees with the impulse model calculations.
In Figs. \ref{fig:2Atdecay} and \ref{fig:3At}, the atoms are initially excited to a singly excited state with the amplitude of the electronic excitation distributed uniformly or to an eigenstate of the complex Green's function matrix of the system. There is no laser interaction and the recoil is measured after the system is allowed to decay into the electronic ground state. Further details are included in Sec. III A of Ref. \cite{sr2021}.

Another inference is that the rate of energy deposited into the system is only dependent on the single atom terms and is not directly dependent on the collective decay dynamics. The single atom term results in the rate of increase of the electronic ground state, and indirectly the rate of accumulation of vibrational excitation, to be proportional to the excitation in the system. 
However, the collective decay dynamics is what determines the lifetime of the excitation. If we integrate the vibrational excitation accumulation over the entire decay process, the energy deposited in such a collective decay will be proportional to the lifetime of the collective excitation. This was also discussed in Sec. III A of Ref. \cite{sr2021}. 

\subsubsection{Longitudinal Oscillation}

In the case of the oscillations in the direction of the separation, the first derivative $g'(\mathbf{R}_{ij})$ in Eq. \eqref{eqn:g_taylor} is no longer non-zero. 
These first order coherent transfers contribute to a $\kappa^2$ order of population transfer. Hence there are two sources of vibrational excitation. Single atom decoherent transfers and first order two atom Lindblad coherent transfers. While the former is unaffected by the trap Hamiltonian, the latter interacts and develops a complicated dependence with the trap Hamiltonian. Figure \ref{fig:2Atdecay}  shows that the recoil in the direction of separation is dependent on the trap frequency and the impulse model is not valid beyond the sudden approximation.

\begin{figure}
    \centering
    \includegraphics[width=0.45\textwidth]{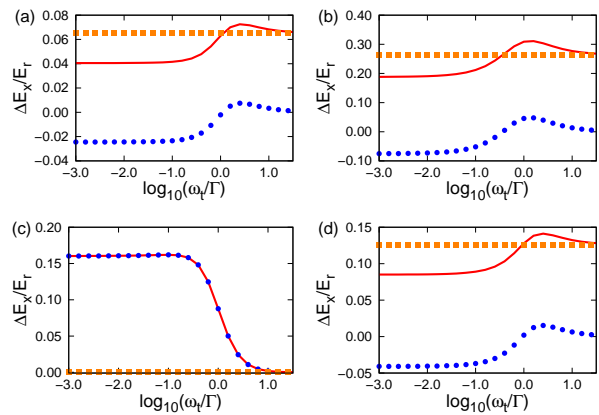}
    \caption{The vibrational energy deposited  during the decay of the excitation. We look at the energy deposited in the x-direction on the center atom when there are three atoms in a line in x-direction separated by $d = 0.4 \lambda$. The red solid line shows the total energy deposited while the blue circles and orange squares show the contribution from the coherent and decoherent transfers respectively.
    The initial excitation is different for the 4 cases. (a) has uniform excitation, (b)-(d) have the 3 eigenstates as excitation. The increase or decrease in energy is dependent on the excitation pattern in the higher $\omega_t$ region. (c) has zero decoherent transfers because the center atom has zero excitation probability in this particular eigenstate. The calculations are done using full density matrix with $\kappa = 0.001$ kept as constant while varying $w_t$.}
    \label{fig:3At}
\end{figure}

Figure \ref{fig:3At} shows an example of the energy deposited in the direction of separation varying with $\omega_t$ when the atoms are initially excited in different distributions. The contributions from the coherent and decoherent transfers are also shown. 
The decoherent transfers are independent of the trap frequency and only depends on the excitation probability of that atom and the decay rate of the system. The coherent transfers, on the other hand, change with the trap frequency and is highly dependent on the the way the excitation is distributed among the atoms and can be either negative or positive. 
The threshold of what determines high trap frequency is set by the collective decay rate of the system and not the individual decay rate of the atom ($\Gamma$). 
While $\kappa$ depends on $M$ and $\omega_t$, in Figs. \ref{fig:2Atdecay} and \ref{fig:3At}, the $\kappa$ is kept as a constant, while altering the trap frequency. This is done in order to isolate the effects of the change in trap frequency while not altering the spread of the wavefunction.

Another distinguishing feature of the coherent and decoherent transfers is the directionality. The coherent transfers are facilitated by the near field dipole-dipole interaction between the two atoms and the recoil in this process is strictly in the direction of separation. The laser interaction is also coherent and has a strict directionality with respect to the direction of the incident light. On the other hand, the decoherent transfer is from spontaneous decay where the direction of photon emission is random and the probability distribution of the direction is governed by the dipole orientation.




\subsection{Comparison with the impulse model\label{sec:comparison}}

The impulse model used in Ref. \cite{sr2021} calculates the kinetic energy and momentum kick imparted in a collective dipole interaction system interacting with a laser. Since the quantum oscillator model discussed in this paper has a fundamental difference in the way the kinetic energy is imparted to the system, the two models can be compared and tested for validity. To account for the spread of the wavefunction, the impulse model can be spatially integrated over the wavefunction probability density using Gaussian quadrature integration for small number of atoms. This result can be compared to the results from the quantum harmonic oscillator model. For low trap frequencies, the results match exactly at low wavefunction spreads and with a small difference for higher spreads. This difference can be shown to be due to stopping at the second order when expanding $g(R)$ in the Taylor's series, i.e., the error is mainly in the harmonic oscillator model. 

The spatial dependence obtained from the impulse model can also be projected on quantum harmonic vibrational states to get the final distribution of the vibrational states. This is an alternate way to validate the two different models. 
In situations where the vibrational states are entangled with the electronic states, the quantum oscillator model might be easier to use than the impulse model. Using the impulse model in this way has not be been explored.


\subsection{Large ensemble of atoms\label{sec:approximations}}    
From Sec. \ref{sec:methods}, the number of states required for calculations increases exponentially with increasing number of atoms. All the atoms having $N_{vib}$ vibrational states would result in all the possible permutation of vibrational state ensembles, i.e., $(N_{vib})^N$ states. 

While the internal state dynamics of absorption, decay, and exchange in excitation are the driving factors of the dynamics of the vibrational states, in the approximation that the spread of the wavefunction is much smaller than the distance separating the atoms, we see that the vibrational state dynamics have little to no effect on the internal state dynamics. Hence we can approximate the calculation so that only one atom is allowed to have quantized vibrational states while the rest are fixed in space. This reduces the total available vibrational states to just $N_{vib}$.  
We calculated the vibrational energy acquired when four atoms in a square are initially excited and decay into the ground state. 
The error when using this approximation is only 0.2\% when the wavefunction spread is as high as 25\% of the separation.

We also see from Sec. \ref{sec:twoatoms} that the second order transfers in the vibrational state are of the order of $\kappa^4$. When taking the expectation of energy, they hardly contribute when $\kappa$ is small. The same reasoning applies to the lasers (as seen in Sec. \ref{sec:singleatom}). Hence we can limit $N_{vib}$ to two without losing generality in this case. For small enough $\kappa = 0.01$, the maximum vibrational energy in the atom can reach up to $10^4 E_r$ which will be within the expected recoil limits. To verify this, the results were tested for convergence using different  $N_{vib}$ in a small number of atoms. 

With these two approximations, we can limit the number of states to $N_{vib} \times (N + 1)$ that is, $ 2(N+1)$ which brings it within the realm of computation for up to 250 atoms.

\subsubsection{Arrays of atoms\label{sec:arrays}}

\begin{figure}
    \centering
    \includegraphics[width=0.45\textwidth]{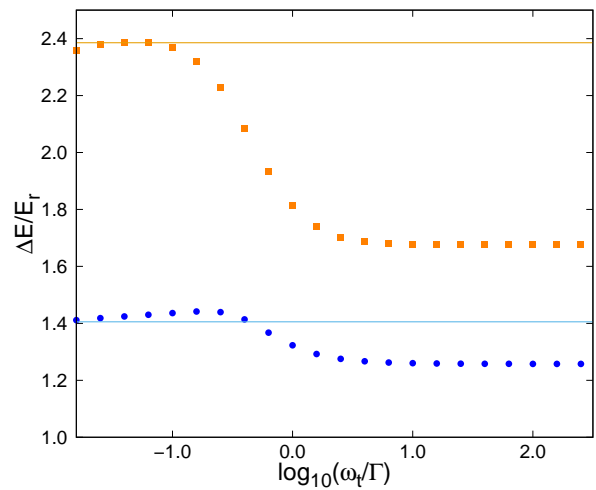}
    \caption{The vibrational energy deposited in the center atom of a $11\times11$ atom array with $d = 0.8 \lambda$ separation when in steady state with an incident laser in the z-direction. The orange squares represent the recoil energy in the z-direction, while the blue circles denote the recoil energy in the x-direction, per photon incident on the center atom. The orange and blue thin lines denote the comparison with the impulse model. This data is calculated using the approximations discussed in Sec \ref{sec:approximations}.}
    \label{fig:25At}
\end{figure}

If there is a constant laser incident perpendicular to an array of closely packed atoms, the recoil in the two different directions have different behaviour.  
Since the laser Hamiltonian does not have two atom interactions, the recoil of the atoms in the direction perpendicular to the array is similar to the single atom laser interaction seen in Sec. \ref{sec:laser}. The recoil within the plane of the array is due to the in-plane collective decay effects as seen in Sec. \ref{sec:twoatoms} and is dependent on the distribution of the excitation. Figure \ref{fig:25At} shows the trend of the recoil in the different directions as a function of the trap frequency. The calculations from the impulse model are also included as a solid line. 




Typically, the trap frequencies in the in-plane (x and y) directions are higher and are about $ 100 $kHz, while the perpendicular trap frequencies are often an order of magnitude lower at about $10 $kHz. These trap frequencies will give a spread of $\kappa / k = 0.08 \lambda $ and $0.025 \lambda$ respectively for a Cs atom. When in steady state, such frequency ranges will be within the slow oscillation approximation and the results from the impulse model can be reproduced with the current model. 

When there is a perfect reflection of a photon from an atom array, there is a momentum of $2 \hbar k$ imparted on the atoms. Hence, the momentum change of the atoms describe the reflectivity of the atom array. This can also be used to study the effects of higher vibrational excited states on the reflectivity. At 10 kHz frequency in the z-direction, the momentum imparted on the central atom of an array reduces by approximately $8\%$ when the atom is in the first vibrational excite state instead of in the ground state. However, at 100 kHz frequency in the z-direction,  there is only a decrease of $0.6\%$. This reinforces that atomic mirror experiments would need to have high trap frequencies to have a reflection probabilities close to 1. 

\subsubsection{Cavity\label{sec:cavity}}
In Ref \cite{sr2021}, we calculated the kinetic energy kick on a cavity, when it decays from a highly subradiant eigenmode. This follows the design of the cavity used in Ref. \cite{zoller} to perform quantum information processing. Under the slow oscillation approximation, the central atom experienced a kick of up to $926 E_r$ in the duration of the decay in the direction perpendicular to the plane. The results were thought to be purely qualitative because of the large lifetimes violating the slow oscillation approximation. 

The results from the Sec. \ref{sec:twoatoms} imply that those calculations were more accurate than suggested in Ref. \cite{sr2021} for the direction perpendicular to the array. In the perpendicular direction, since there are no or negligible inter-atom forces, the trap frequency does not play a significant role in determining the vibrational energy deposited. 
The recoil due the decoherent transfers accumulate over an extended duration due to the subradiant decay resulting in large recoil energies deposited.
Another way to interpret this is the large quality factor causing there to be multiple reflections of the photon on the array faces. 

Calculations for the same cavity as that in Ref. \cite{sr2021}, using the harmonic oscillator model resulted in the center atom experiencing similar vibrational energy deposited, approximately $922 E_r$ in the direction perpendicular to the array. This recoil was unaffected when the trap frequencies were increased beyond the decay rate of the system. 
On the other hand, the energy deposited in the in-plane direction at high frequencies decreased to $15.0 E_r$ as compared to $16.6 E_r$ at low frequencies.
These results show that the recoil of the atoms due to collective decay, especially in highly subradiant systems, should not be ignored.

\section{Conclusion\label{sec:conclusion}}

We presented a model to describe and calculate the recoil in light-matter collective interaction using quantum harmonic oscillator trap potentials. 
We compare the results of the impulse model used in Ref. \cite{sr2021} under the slow oscillation approximation and explored the regime beyond. We studied the contribution to recoil from the different terms of the Hamiltonian and Lindblad equation. In essence, the single-atom Lindblad term causes a recoil in a random direction and the energy deposited is independent of the trap frequency used. The laser Hamiltonian causes a recoil in the direction of the laser propagation and recoil energy deposited falls off to zero when the trap frequency goes beyond the collective decay rate of the system. The two-atom Lindblad terms induce a recoil in the direction of the separation between the atoms and it is dependent on both the trap frequency and the distribution of the excitation in the system.

In atom arrays, in the directions where there are no inter-atom forces or lasers, the recoil is independent of trap frequency and the impulse model can be used even beyond the slow oscillation approximation.
If the atoms are excited by a laser or for those directions in the plane of the atom array, the impulse model is no longer valid when the trap frequency approaches or is higher than the decay rate of the system.

This model was used to verify the extremely high recoil calculated in a cavity with high subradiance. This shows that recoil effects have to be considered seriously when working with highly subradiant systems. The effects of vibrational excitation in the reflectivity of arrays were also studied. 

References \cite{shahmoon,sly2019} have worked on the opposite regime of the sudden approximation, where the focus lies on the slow center of mass motion rather than the fast internal state dynamics. Studying this regime, especially the collective modes of vibration of the atoms using the quantum harmonic oscillator model could lead to better understanding and control of atom arrays. 

Data used in this publication is available at \cite{datalink}.

\begin{acknowledgments}
This work was supported by the National Science Foundation under Award No. 2109987-PHY. This research was supported in part through computational resources provided by Information Technology at Purdue University, West Lafayette, Indiana.

\end{acknowledgments}
 
 \bibliography{ref.bib}
 
\end{document}